\documentclass[conference]{IEEEtran}
\IEEEoverridecommandlockouts
\usepackage{cite}
\usepackage{amsmath,amssymb,amsfonts}
\usepackage{algorithmic}
\usepackage{graphicx}
\usepackage{float}
\usepackage{multirow}
\usepackage{textcomp}
\usepackage{booktabs}
\usepackage{multirow}
\usepackage{array}
\usepackage{xcolor}
\def\BibTeX{{\rm B\kern-.05em{\sc i\kern-.025em b}\kern-.08em
    T\kern-.1667em\lower.7ex\hbox{E}\kern-.125emX}}
\begin{document}

\title{FL-DABE-BC: A Privacy-Enhanced, Decentralized Authentication, and Secure Communication for Federated Learning Framework with Decentralized Attribute-Based Encryption and Blockchain for IoT Scenarios \\
}

\author{
    \IEEEauthorblockN{
        Sathwik Narkedimilli\IEEEauthorrefmark{1},
        Amballa Venkata Sriram\IEEEauthorrefmark{1},
        Satvik Raghav\IEEEauthorrefmark{2}
    }
    \IEEEauthorblockA{
        \IEEEauthorrefmark{1}Department of Computer Science, Indian Institute of Information Technology (IIIT) Dharwad, Dharwad\\
        Email: 21bcs103@iiitdwd.ac.in, 21bcs008@iiitdwd.ac.in
    }
    \IEEEauthorblockA{
        \IEEEauthorrefmark{2}Department of Electronics and Communication Engineering, Amrita Vishwa Vidyapeetham Bengaluru\\
        Email: satvikraghav007@gmail.com
    }
}

\maketitle

\begin{abstract}
In IoT scenarios, where data privacy and security are among the major concerns, FL (Federated Learning) is a decentralized training approach that can be used to train machine learning models without transferring sensitive data from IoT devices. The framework proposed in our research advances the concept of FL for IoT scenarios, adding the most advanced security schemes such as Decentralized Attribute-Based Encryption (DABE) for decentralized authentication, Homomorphic Encryption (HE), Secure Multi-Party Computation (SMPC), and Blockchain technology to secure the entire process of modeling training against data privacy and security issues. In this framework, data from all IoT devices are locally encrypted using DABE for decentralized authentication and data encryption, then based on HE, data can be securely computed on encrypted data. Afterward, SMPC technology is employed to preserve privacy and collaborative computations. The output from SMPC is securely transmitted via Blockchain, which guarantees transparent communication and data integrity and securely stores all relevant transactions and model updates in a distributed ledger across the entire FL network.

The framework begins with IoT devices collecting and preparing data for local model training, where the data is encrypted using DABE. Initial deep learning models configured in cloud servers are distributed to edge devices via a blockchain network, ensuring immutable record-keeping and secure peer authentication. After local training, the updated model weights are encrypted and securely transmitted to fog layers via the blockchain, where micro-services perform aggregation using homomorphic encryption and SMPC. The FL server aggregates these local updates with differential privacy to prevent data leakage and iteratively refines the global model. The final global model is then distributed back to the IoT devices for deployment, enabling real-time analytics in IoT market spaces. This framework effectively addresses the challenges of secure decentralised learning in IoT environments, offering a robust solution for privacy-preserving, efficient, and secure federated learning.
\end{abstract}


\section{Introduction}

The Internet of Things (IoT) \cite{article4985975} is a brilliant way to connect different physical objects with digital information. Its rapid adoption is already revolutionizing many different aspects of different industries. IoT has already resulted in amazingly future-oriented applications, including consumer-grade home automation and medical science. But if we want to see further growth in these areas and move to new areas such as industrial automation, we also need to address the key challenges of addressing data privacy and security concerns. Centralized machine learning models, often involving the transfer of sensitive information from IoT devices to centralized servers, have proven to be inadequate for these purposes. They also potentially expose data to a variety of malicious attacks. This has made it an increasingly urgent priority to move to decentralized, secure learning frameworks.

FL has the potential to overcome these issues as it enables the training of machine learning models on devices without transferring the underlying data into a central server. This is facilitated by the FL approach of allowing trained models to be updated individually on IoT devices and shared with a central server as part of the iterative aggregation process. However, a lack of security in FL frameworks could still result in privacy leaks and data exposure in IoT environments, which process sensitive data. This makes it crucial to develop robust security strategies that can be seamlessly incorporated into FL frameworks to achieve a high level of data privacy and integrity, such as using distributed data.

Our proposed framework named FL-DABE-BC, which is a novel privacy-pres framework for IoT environments, combines key components of Decentralized Attribute-Based Encryption (DABE), Homomorphic Encryption (HE), Secure Multi-Party Computation (SMPC), and Blockchain. FL-DABE-BC first uses DABE for decentralized authentication and fine-grained data encryption on IoT devices. The data encryption with DABE is done on the local IoT device, thus reducing the chances of a breach of sensitive information and enhancing the security of the overall federated learning process. Additionally, the integration of Microservices \cite{BENATITALLAH2023100845} allows for scalable and modular deployment, enabling the framework to efficiently manage and process IoT data while maintaining high levels of security and privacy.

Enhancing data privacy and secure computation capabilities, FL-DABE-BC leverages Homomorphic Encryption and Secure Multi-Party Computation. Homomorphic Encryption enables complex computations to be performed directly on encrypted data. By leveraging this technique, data doesn’t need to be exposed and available for easier processing. It assures data privacy during aggregation, processing, and model training as well. At the same time, SMPC enables the parties to cooperate to perform computations on their private inputs while keeping them confidential to other parties involved in the process. This method adds serious robustness to data privacy and security since it mitigates vulnerabilities in distributed learning.

To ensure transparency and immutability in communication between devices/servers in the FL-DABE-BC framework, blockchain technology is employed as the communication channel for the transfer of model updates and the integrity of the data. All the transaction and model update information deployed on the blockchain is stored in the distributed ledger in a secure manner, which is impossible to be modified by a single device/server. Thus, the collaboration among the participating devices/servers in federated learning can be enhanced with this transparent and secure distributed ledger system. With the integration of blockchain, the FL-DABE-BC framework proposed here can provide an end-to-end solution to IoT-based federated learning, which addresses the issues of privacy, security, integrity, and effective communication between the nodes (EDGE-FOG-CLOUD layers). The integration of peer-to-peer (P2P) networks like blockchain has proven to improve the efficiency of communication by optimizing bandwidth \cite{8875437} and ensuring secure data exchange. As a result, the FL-DABE-BC framework offers an end-to-end solution for IoT-based federated learning that addresses privacy, security, integrity, and effective communication across the EDGE-FOG-CLOUD layers.

\section{Preliminaries}

\subsection{Federated Learning}

Federated Learning (FL) \cite{ZHANG2021106775} is the machine-learning paradigm that provides a decentralized approach to train models on end-user devices such as smartphones or Internet-of-Things (IoT) devices without having to move the raw data to a central server. In this way, the data stays on the user devices, and only model updates like gradients or weights are shared with the central server for combining (aggregation). FL is an important step towards minimizing risks of data privacy and security concerns by removing the need for centralized data storage and processing, which is becoming commonplace as our society increasingly moves services into the cloud. It also takes full advantage of the computational power of edge devices. As such, FL is very useful for scenarios where the data is sensitive, especially when the network bandwidth is limited.

\subsection{Decentralized Attribute-Based Encryption}

DABE (Decentralized Attribute-Based Encryption) \cite{lewko2011decentralizing} is a cryptographic technique that enhances security and privacy by allowing data to be encrypted based on specific user attributes rather than a single key or identity. In a decentralized setting, such as in IoT environments, DABE enables fine-grained access control and secure data sharing among multiple users and devices. Unlike traditional encryption methods, where a central authority manages encryption keys, DABE distributes the control among multiple authorities, reducing the risk of a single point of failure. This decentralized approach aligns well with federated learning, as it allows data to be securely encrypted on each device based on user attributes, ensuring that only authorized entities can decrypt and access the data.

\subsection{Blockchain}

Blockchain \cite{zheng2018blockchain} is a distributed, immutable ledger system able to record transactions in a secure, transparent, and tamper-proof way. In federated learning, it can be used to verify and log updates to models, with all transactions openly broadcasted, recorded according to a timestamp, and accessible to all participants. It can be used for secure peer-to-peer communication between IoT devices and fog nodes, as well as for decentralized authentication and executing privacy-preserving operations without exposing the original data or models to unauthorized entities who might try to modify the data or model transactions. In this way, it can provide a trustworthy environment to execute collaborative learning without centralized supervision and oversight.

\subsection{Cryptographic Primitives}

\begin{enumerate}

\item \textit {Homomorphic Encryption (HE)}: It \cite{yi2014homomorphic} enables computations to be carried out directly on encrypted data in such a way that the results are still encrypted; as a result, no sensitive information needs to be revealed during the processing stages. 

\item \textit {Secure Multi-Party Computation (SMPC)}: It \cite{cramer2015secure} allows multiple parties to jointly compute a function on their inputs while keeping their inputs private. The second application is federated learning. In federated learning, multiple devices or users collaborate in training a model while keeping their data private.

\item \textit {Differential Privacy (DP)}: It \cite{dwork2006differential} is a technique that, by introducing controlled noise in the data or the model outputs, prevents the identification of individual data points even if the aggregate data is revealed.

\end{enumerate}

These cryptographic primitives provide strong privacy and security guarantees; they are the essential building blocks for designing federated learning frameworks that work in IoT environments.
 
\section{Literature Review}

The paper ‘Recent Advances on Federated Learning: A Systematic Survey’ \cite{LIU2024128019} reviews seminal FL frameworks. FedAvg is the initial algorithm that averages the client model updates offline and allows decentralized learning. At the same time, FedMA aligns and averages parameters from clients with heterogeneous data to tackle heterogeneous data challenges. FedProx extends FedAvg, adding a proximal term to address the heterogeneity of client data and achieve stable model convergence. The algorithms in this paper are among the initial developments for FL, from simple aggregation to more complicated methods in dealing with client diversity. The other paper introduces FedSGD \cite{konečný2017federatedlearningstrategiesimproving}, which adopts synchronous updates to ensure model consistency but suffers from communication overhead and stragglers.

The application of federated learning (FL) to Intrusion Detection Systems (IDS) \cite{AGRAWAL2022346} presents a promising approach for enhancing data privacy and scalability in real-time detection scenarios. Research highlights the integration of distributed IDS to manage large-scale network data while preserving user privacy. However, challenges such as communication overhead, data source heterogeneity, and incorporating advanced threat intelligence within FL models remain significant. These issues underscore the need for optimized frameworks that balance privacy, efficiency, and effective threat detection in diverse and complex environments.

In intelligent transportation systems (ITS), federated learning frameworks have been adapted to address the dynamic nature of vehicular networks. The V2X-Boosted FL model, incorporating contextual client selection \cite{unknown45346346}, prioritizes relevant and high-quality data for training, optimizing resource use and model accuracy. Similarly, the semi-asynchronous hierarchical FL framework \cite{unknown425692365126} accommodates varying data processing capabilities among vehicles, reducing communication latency and enhancing real-time applicability. Both approaches highlight the ongoing challenges of data heterogeneity, dynamic client availability, and secure communication within ITS.

To enhance privacy in federated learning, combining group signatures and FL \cite{KANCHAN202393} offers a robust solution, as demonstrated in recent studies. This method allows for the verification of data authenticity without revealing participant identities, thus preserving privacy and trust in decentralized networks. Additionally, the integration of blockchain and smart contracts with FL \cite{article937593757} introduces a tamper-proof ledger for model updates, addressing security and transparency issues. Despite these advancements, challenges such as computational overhead, scalability, and smart contract deployment complexities continue to demand innovative solutions.

Incorporating local differential privacy (LDP) within federated learning framework \cite{BATOOL2024119717} is a growing area of interest, particularly in sensitive applications such as Power IoT systems and Vehicular Ad Hoc Networks (VANETs). The IFed framework \cite{article34o039460374067} and FL with LDP for VANETs both aim to maintain data privacy while allowing real-time data processing. These models effectively mitigate privacy risks and protect against inference attacks. However, they face challenges like resource constraints, data sparsity, and maintaining model accuracy without compromising privacy, highlighting areas for further research and development.

Recent advancements in federated learning (FL) and secure communication have improved IoT data analytics and community systems. FL frameworks like FedMicro-IDA \cite{BENATITALLAH2023100845} and V2X-boosted FL optimizes privacy, efficiency, and model accuracy, despite challenges like dynamic client availability. Blockchain and smart contracts enhance security with tamper-proof ledgers, while local differential privacy (LDP) ensures secure data sharing \cite{SHARMA2021102790}. Scalability and resource constraints, however, remain significant challenges in these systems.

\section{Proposed Model}

\subsection{The FL-DABE-BC Workflow}

The FL-DABE-BC framework uses Decentralized Attribute-Based Encryption (DABE) for secure data encryption and blockchain for immutable records in IoT environments. Data is encrypted locally, aggregated using homomorphic encryption and SMPC, and refined with differential privacy. The final global model is securely deployed for real-time analytics. Fig 1 demonstrates the workflow of the model and the steps include : 

\begin{enumerate}

\item \textit {Data Collection and Preparation by IoT Devices}:

The process starts with IoT devices in the market spaces such as retail stores and smart cities collecting raw data from their surroundings in the form of customer interactions, environmental conditions, and sensor-derived metrics. Then, the raw data is preprocessed at the edge of each IoT device to make it ready for model training. This is done by normalizing the data to achieve a consistent aspect and extracting features to highlight the important patterns. Further, labeling these data points is done to categorize them.

Anonymized, preprocessed data is then encrypted using a technology known as Decentralized Attribute-Based Encryption (DABE) before it can be accessed. DABE only permits the decryption of data based on defined attributes, enabling only the fog nodes or cloud servers with the correct credentials to see, use, or analyze the data. This prevents any unauthorized entities from gaining access to it, which is important for ensuring sensitive data’s privacy at the edge, where IoT devices operate. \\

\item \textit {Initial Model Configuration and Distribution}:

After the data has been properly prepared, the cloud servers only initiate the work of training initial DL models, which would be customized based on the nature of the available data and the corresponding machine-learning tasks. To ensure the model’s integrity during transmission, homomorphic encryption would be performed on the initial models. Homomorphic encryption allows computations to be performed on the encrypted data, meaning that the models can be securely transmitted over the network.

Encrypted model copies are provided to edge clients (the IoT) via a blockchain network. A blockchain maintains unchangeable and transparent records of all model transactions. The immutability of blockchain transactions enhances security since it makes tampering or unauthorized changes impossible. Every IoT device needs to verify the identity of their peers using DABE, before initiating the download of the model. Only authenticated peers can engage in the training process. \\

\item \textit {Local Training on IoT Devices}:

Once the models are in place, local training with encrypted IoT data can begin: every device can train on its encrypted data in a distributed manner. Each device can do its part to update the model based on its data. This procedure is repeated multiple times so that the model improves and ‘fine-tunes’ depending on the specific characteristics of the local data.

After each iteration, the locally updated model weights are encrypted using DABE. This step ensures that the locally updated model weights cannot be compromised through eavesdropping, or be accessed by unauthorized nodes before they are stored or transmitted to other nodes. More importantly, only authorized nodes (such as fog nodes or cloud servers) can decrypt and utilize the updated weights. \\ 

\item \textit {Communication and Secure Aggregation}:

It’s the encrypted local model weights that get uploaded to the fog layer via the blockchain network. The aggregation of these weights is performed by microservices deployed within the fog layer. Since the weights need to be aggregated by the fog nodes, this needs to happen under conditions that maintain data confidentiality. Homomorphic encryption is employed to achieve this: it allows the fog nodes to combine the weight values without the need to decode the individual data elements first.

Security and privacy are further enhanced by using Sec Computation (SMPC) between the nodes in the fog layer. SMPC allows multiple parties to jointly compute aggregate weights without revealing the inputs that each party used in the computation. Each participating node contributes to the computation but its local data remains private. Blockchain is used to securely log and track the SMPC computations in a transparent and tamper-proof manner. \\

\item \textit {Aggregating Local Model Weights}:

The FL server in the cloud collects the aggregated model weights from the fog layer and, using homomorphic encryption, it can then aggregate these weights without ever having to decrypt them. This way the FL server will not know what each node is contributing, and the model weights in the aggregated model will never be exposed to the cloud as they travel as encrypted data. The data stays confidential, and the integrity of the model-training process is maintained.

Once the weights are combined, the model is subjected to differential privacy techniques. Differential privacy ‘whitens’ the combined model, preventing leakage of individual data points, memorization of data, and protection of overall model privacy. This phenomenon is key to both protecting the privacy of users and enabling the model to learn from the diversity of available datasets. \\

\item \textit {Iterative Process of Model Training}:

The federated learning process is composed of multiple iterative communication rounds. In each round, the IoT devices carry out local training, encrypt and upload their latest weights to the cloud, and take part in the federated aggregation implemented by the fog layer and the blockchain. The FL server aggregates the model weights and updates the global model. The blockchain technology guarantees that every message communication between the nodes is safe and traceable, making it impossible for any adversary to access or tamper with these channels.

At the end of every aggregation, the new world model is encrypted and sent back to the FL clients (IoT devices and fog nodes) through the blockchain network. The FL clients authorize their server via DABE and update their local models with the new world model. This process is repeated until the world model achieves certain performance metrics (eg, accuracy or loss), as determined by predefined thresholds. \\

\item \textit {Final Model Deployment and Local Data Analytics}:

Once the global model achieves effective performance, it is finalized, encrypted, and authenticated through DABE. The final global model is then deployed to all IoT devices and fog nodes, allowing them to use the model for real-time data analytics. IoT devices utilize the updated global model to make informed decisions and generate insights based on new data collected from their environments, such as customer behavior analysis or inventory management. \\

\item \textit {Microservices in the Fog Layer}:

Microservices are deployed on fog nodes to manage various tasks, including data aggregation, real-time analytics, and secure communication within the blockchain network. These microservices provide a modular and scalable architecture, allowing the framework to efficiently process data and perform computations at the network's edge. This design ensures efficient local processing, reducing latency and improving overall performance in large-scale IoT deployments. 

\end{enumerate}

\begin{figure}[htp]
    \centering
    \includegraphics[width=9.0cm, height=8.6cm]{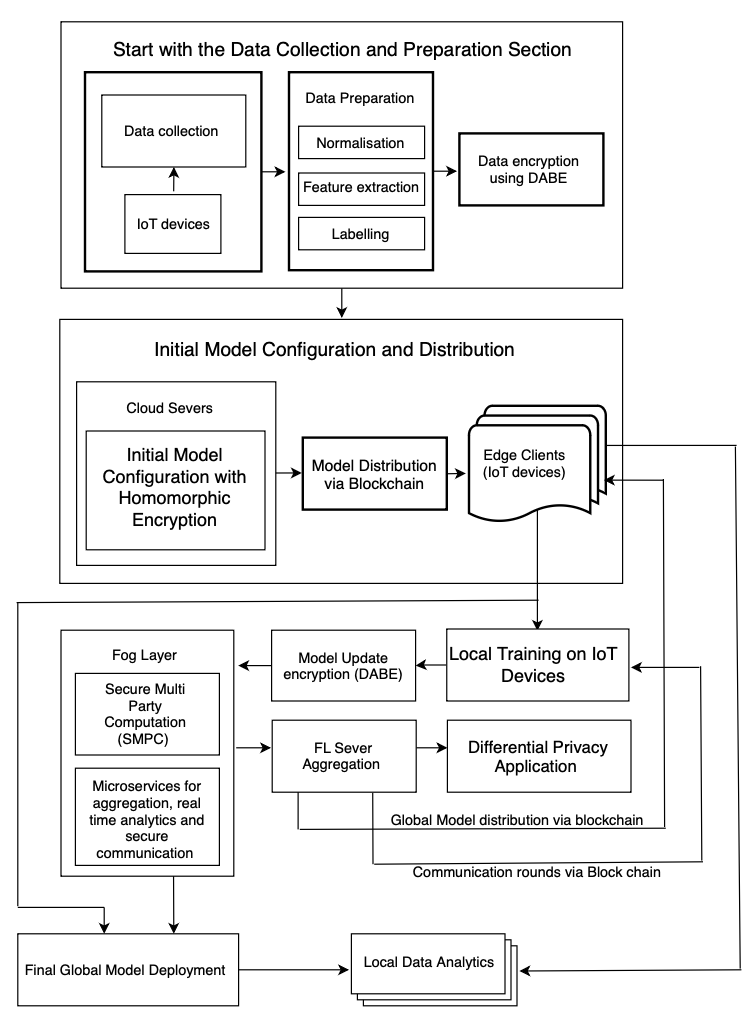}
    \caption{Architecture of the proposed model}
    \label{fig:architecture}
\end{figure}

\subsection{Assumptions for FL-DABE-BC}

There are several assumptions underlying the FL-DABE-BC framework. Every entity (IoT device, fog node, microservices, or cloud server) in the framework assumes that all IoT devices, fog nodes, microservices, and cloud servers have a unique cryptographic key managed by the DABE system. The DABE system will ensure that the key for each user has attributes that are granted to every entity whose attributes match. It is assumed that all entities trust the DABE system. They assume that the system is correctly implemented and secure so that nobody can tamper with it or access a key without the correct attributes. Third, the framework assumes the secure implementation of homomorphic encryption and SMPC that allows aggregating encrypted data without exposing sensitive information.

Furthermore, the underlying blockchain network is presumed to be secure and immutable, offering open and auditable logging of all transactions, data exchanges, and updates of models. The blockchain is trusted to manage and record all SMPC processes, ensuring correctness and transparency for all federated learning activities. The integrity of the blockchain ensures that all IoT devices, fog nodes, microservices, and clouds in the showcase IoT market spaces can trust activities and exchanges. The differential privacy techniques are assumed to prevent data leakage, keeping the privacy of each data point in place even after aggregation. These assumptions collectively allow for the federated learning framework to securely work in IoT market spaces, maintaining data integrity and privacy.

The proposed federated learning framework is scalable, can work in dynamic environments, and can leverage the increasing number of IoT devices and fog nodes. Furthermore, DABE and blockchain transparently and securely manage data in a decentralized manner across multiple computing nodes. The microservices in the fog layer support scalable modular data processing and also allow tasks to be parallelized. Blockchain (with a layer-2 solution like offloading computation to other nodes using zkML \cite{so2024oppaioptimisticprivacypreservingai}) enables seamless communication and integration, facilitating the federated learning framework to scale out over multiple nodes to meet large data storage and computational needs in large IoT deployments.

\subsection{BAN Logic analysis for FL-DABE-BC}

BAN Logic (Burrows-Abadi-Needham Logic) \cite{inproceedings4665} is a formal method for reasoning about authentication protocols. It employs a set of rules and notational conventions to model the reasoning of the parties in a communication protocol. BAN logic helps to uncover flaws: weak or ambiguous beliefs can be quickly identified. Its task can be summarised as explicitly representing what each party believes about identities, message freshness, and trust in the data they’re receiving. This way, protocols can be checked against the desired security properties.

The following is the BAN logic for the FL-DABE-BC framework, analyzing its authentication and security properties. By using BAN logic, the framework’s assumptions and trust relationships are clearly defined, ensuring that its protocols are secure and correctly implemented.

\subsection*{Notations in BAN Logic:}
\begin{itemize}
    \item $P \rhd X$: \textbf{Principal} $P$ \textbf{believes} statement $X$ is true.
    \item $P \triangleleft X$: \textbf{Principal} $P$ \textbf{has received} statement $X$.
    \item $P \stackrel{\text{says}}{X}$: \textbf{Principal} $P$ \textbf{at some point sent} statement $X$.
    \item $P \rhd \# X$: \textbf{Principal} $P$ \textbf{has jurisdiction over} statement $X$.
    \item $\{X\}_K$: \textbf{The message} $X$ \textbf{is encrypted with key} $K$.
    \item $K \leftrightarrow P$: \textbf{K is a shared key between} $P$ \textbf{and another principal, known only to them}.
\end{itemize}

\subsection*{Entities in the Framework:}
\begin{itemize}
    \item IoT devices $(D)$
    \item Fog nodes $(F)$
    \item Microservices in Fog Layer $(M)$
    \item Cloud servers $(C)$
    \item Blockchain network $(B)$
\end{itemize}

\subsection*{Assumptions and Initial Beliefs:}
\begin{itemize}
    \item \textbf{A1:} All entities believe in the validity of their cryptographic keys:
    \begin{align*}
        D &\rhd (K_{D, F} \leftrightarrow F) \\
        F &\rhd (K_{F, M} \leftrightarrow M) \\
        M &\rhd (K_{M, C} \leftrightarrow C) \\
        C &\rhd (K_{C, B} \leftrightarrow B)
    \end{align*}

    \item \textbf{A2:} All entities believe that blockchain transactions are securely logged and can be trusted:
    \begin{align*}
        D &\rhd (B \rhd \# \text{Transactions}) \\
        F &\rhd (B \rhd \# \text{Transactions}) \\
        M &\rhd (B \rhd \# \text{Transactions}) \\
        C &\rhd (B \rhd \# \text{Transactions})
    \end{align*}

    \item \textbf{A3:} All entities believe in the correctness and security of Decentralized Attribute-Based Encryption (DABE), Homomorphic Encryption (HE), and Secure Multi-Party Computation (SMPC):
    \begin{align*}
        D &\rhd \# \text{DABE is secure} \\
        F &\rhd \# \text{DABE is secure} \\
        M &\rhd \# \text{HE and SMPC are secure} \\
        C &\rhd \# \text{HE is secure} \\
        B &\rhd \# \text{HE and SMPC are secure}
    \end{align*}

    \item \textbf{A4:} Fog nodes $(F)$ believe in the integrity and secure operation of microservices $(M)$ for aggregating model updates:
    \begin{align*}
        F &\rhd M \rhd \# \text{Aggregation and Computation}
    \end{align*}
\end{itemize}

\subsection*{Goals of Authentication and Security:}
\begin{itemize}
    \item \textbf{G1:} IoT devices $(D)$ believe in the integrity of the model updates received from the cloud servers $(C)$:
    \begin{align*}
        D &\rhd C \rhd \text{Global Model}
    \end{align*}

    \item \textbf{G2:} Fog nodes $(F)$ and microservices $(M)$ believe in the integrity of data and model updates sent from IoT devices $(D)$:
    \begin{align*}
        F &\rhd D \rhd \text{Local Model Updates} \\
        M &\rhd F \rhd \text{Aggregated Model Updates}
    \end{align*}

    \item \textbf{G3:} Cloud servers $(C)$ trust the secure aggregation of model updates performed using SMPC over blockchain $(B)$:
    \begin{align*}
        C &\rhd \# (B \rhd \text{SMPC Aggregation})
    \end{align*}
\end{itemize}

\subsection*{Protocol Messages Using BAN Logic:}
\begin{itemize}
    \item \textbf{M1:} IoT devices $(D)$ send encrypted data to fog nodes $(F)$ using DABE and log the transaction on blockchain $(B)$:
    \begin{align*}
        D \rightarrow F: \{\text{Data}\}_{K_{D, F}}, \text{log}(\text{Data Transfer})_B
    \end{align*}
    \textbf{Postulate:} \( F \triangleleft \{\text{Data}\}_{K_{D, F}} \)

    \item \textbf{M2:} Fog nodes $(F)$ forward encrypted local models to microservices $(M)$ for aggregation, and log the process on blockchain $(B)$:
    \begin{align*}
        F \rightarrow M: \{\text{Local Model}\}_{K_{F, M}}, \text{log}(\text{Forwarding})_B
    \end{align*}
    \textbf{Postulate:} \( M \triangleleft \{\text{Local Model}\}_{K_{F, M}} \)

    \item \textbf{M3:} Microservices $(M)$ aggregate the local models using homomorphic encryption and SMPC, and send aggregated models to cloud servers $(C)$, logging the operation on blockchain $(B)$:
    \begin{align*}
        M \rightarrow C: \{\text{Aggregated Model}\}_{HE}, \text{log}(\text{Aggregation})_B
    \end{align*}
    \textbf{Postulate:} \( C \triangleleft \{\text{Aggregated Model}\}_{HE} \)

    \item \textbf{M4:} Cloud servers $(C)$ update the global model and broadcast the encrypted global model updates back to IoT devices $(D)$ via blockchain $(B)$:
    \begin{align*}
        C \rightarrow D: \{\text{Global Model Update}\}_{K_{C, D}}, \text{log}(\text{Update})_B
    \end{align*}
    \textbf{Postulate:} \( D \triangleleft \{\text{Global Model Update}\}_{K_{C, D}} \)
\end{itemize}

\subsection*{Derivations and Authentication Verification:}
\begin{itemize}
    \item \textbf{D1:} From \textbf{M1}, \textbf{A1}, and \textbf{A3}:
    \begin{align*}
        F &\rhd (D \rhd \text{Data})
    \end{align*}
    Fog nodes believe the data sent by IoT devices is authentic and securely encrypted.
    
    \item \textbf{D2:} From \textbf{M2}, \textbf{A1}, \textbf{A2}, and \textbf{A4}:
    \begin{align*}
        M &\rhd (F \rhd \text{Local Model Updates})
    \end{align*}
    Microservices believe the local model updates from fog nodes are authentic and securely encrypted.
    
    \item \textbf{D3:} From \textbf{M3}, \textbf{A1}, \textbf{A2}, \textbf{A3}, and \textbf{A4}:
    \begin{align*}
        C &\rhd (M \rhd \text{Aggregated Model})
    \end{align*}
    Cloud servers believe the aggregated model updates from microservices are authentic, securely computed, and encrypted.

    \item \textbf{D4:} From \textbf{M4}, \textbf{A1}, \textbf{A2}, and \textbf{A3}:
    \begin{align*}
        D &\rhd (C \rhd \text{Global Model Update})
    \end{align*}
    IoT devices believe the global model updates from cloud servers are authentic and securely transmitted.\\
\end{itemize}

Based on the foregoing explanation, when we make a BAN logic analysis, it is clear that the federated learning architecture applied to the IoT scenarios with microservice-based architecture and blockchain can provide a secure, robust, and privacy-preserving environment for model training. DABE, HE, and SMPC can be applied for authentication and integrity proof during data and model updates between IoT devices, fog nodes, microservices, and cloud servers, while the blockchain network can be used for trustworthy logging, leading to provable auditability of entire transactions. This combination of the technologies guarantees that both the entire learning process and the privacy of data can be provided securely to cater to reliable decentralized learning and gain the trust of all participants.

\section{Analysis of Proposed Model}

Our proposed federated learning framework utilizes blockchain technology and sophisticated cryptographic primitives such as Decentralised Attribute-Based Encryption (DABE), Homomorphic encryption, and Secure Multi-Party Computation (SMPC) to enhance data privacy and security during collaborative learning, from data contribution to model transmission and aggregation. The use of DABE can ensure that only authorized fog nodes and cloud servers with the right attributes can access and decrypt the data, thus preventing unauthorized access. It helps protect from data privacy attacks such as data poisoning, etc Homomorphic encryption can secure the model transmission and model aggregation, ensuring that computations on the data can be made without revealing the original data. SMPC can keep the raw data private during model aggregation by allowing encrypted exchanges between fog nodes and cloud servers.

Moreover, it resolves security issues inherent to centralized models by spreading the tasks across multiple entities using blockchain technology – a tamper-proof ledger that stores an immutable history of any transactions, model updates, and computations, thereby making it more transparent, secure and reducing a single point of failure. Meanwhile, the use of differential privacy in aggregating model weights ensures that the data would not leak and avoids data memorization, protecting users’ privacy and preserving the accuracy of the model. These components seamlessly work together to create a robust federated and secure learning environment to counter the various data privacy and security threats present in IoT scenarios. FL-DABE-BC is a robust framework ensuring data privacy, integrity, complete transparency, and accountability in the system and ensuring secure and effective communication between the nodes. It is a trustworthy, reliable, and secure framework for collaborative learning. 

The following analysis and (Fig. 2) details how FL-DABE-BC holds up against various vulnerabilities and attacks:

\subsection{Anonymity}

Anonymity is built in the federated learning framework using the Decentralized Attribute-Based Encryption (DABE). In this framework, data coming from IoT devices is encrypted with DABE. Those that can decrypt the data are only those with the right attributes. Thus, any participant (IoT devices as data sources) is anonymous to any outsider (attacker) who has no right attribute. Blockchain also guarantees anonymity in recording all transactions without revealing the identities of the blockchain network and can engage in pseudonymous transactions.

\subsection{Non-Traceable and Impersonation Attacks}

Furthermore, the framework can also mitigate non-traceable and impersonation attacks through the use of Decentralised Attribute-Based Encryption (DABE) and blockchain. DABE will ensure that only users who hold the proper credentials can study data or be part of the distributed network, thus preventing impersonation or unauthorized access. Moreover, due to the inherent nature of federated learning, that is, staying on-device, the data source is untraceable, thus preventing data tracing. And, through blockchain’s inherent immutable ledger, it becomes impossible for an attacker to impersonate a legitimate node without being detected. Moreover, the pure use of blockchain technology by itself guarantees that attackers will not be able to trace transactions back to their source as due to the use of secure cryptographic hashes, the whole blockchain ledger is protected.

\subsection{Message Modification Attacks}

The proposed framework protects against message modification attacks by leveraging homomorphic encryption (HE) and blockchain. With homomorphic encryption, model weights, and data remain encrypted throughout the training and aggregation processes, ensuring that even if an attacker intercepts the messages, they cannot modify the content without detection. Furthermore, the blockchain records every transaction in a tamper-proof ledger. Any attempt to modify a message or model update would require altering all subsequent blocks, which is computationally infeasible. This ensures the integrity of messages exchanged between IoT devices, fog nodes, and cloud servers.

\subsection{Replay Attacks and Man in the Middle Attacks}

The framework protects against replay and man-in-the-middle attacks using blockchain and SMPC. Because the ledger provided by the blockchain is unchangeable, the blockchain prevents replay attacks by establishing that every transaction is independently unique and cannot be copied without this uniqueness being detectable and unacceptable. Since the blockchain provides a secure communication channel between the clients that are generating the data to be aggregated, any attempts at man-in are detectable as the inputs must pass cryptographic checking before entering the blockchain and hence entering the aggregation protocol. Furthermore, the inputs are ‘blinded’ by the SMPC aggregation protocol so that, if an attacker were to capture the data during the aggregation, nothing useful could be discerned about the inputs so the computation process retains its confidentiality and integrity.

\section{Conclusion}

This work proposes a federated learning (FL) framework for large-scale IoT environments to provide data security, privacy, and decentralized authentication using blockchain, DABE, HE, SMPC, and differential privacy. FL utilizes DABE to limit who can access the encrypted data and provides a way to control access. Encrypted data can then be securely aggregated or computed using HE and SMPC without revealing individual data points to a third party. Blockchain provides transparency, immutability, and trust in all transactions and model updates as well as secure and efficient communication between the network. This model provides security, decentralization, and privacy-preserving FL for large-scale IoT ecosystems and can be scaled for large-scale implementations with several microservices and users.
Most importantly, this framework is efficient regarding time and computational overhead. Differential privacy also ensures that the individual data contributions remain private during aggregation. This work presents an innovative and promising approach but still has broad and exciting areas for future work.

\begin{figure*}[h]
    \centering
    \includegraphics[width=\textwidth, height=8.5cm, keepaspectratio]{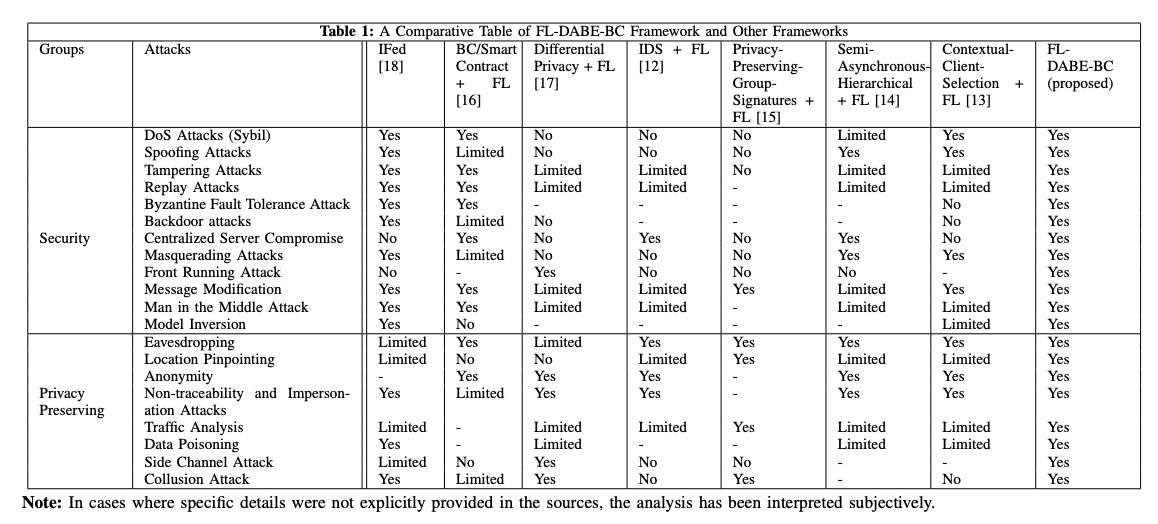}
    \caption{Comparision Table with different Federated Learning Frameworks}
    \label{Table: Comparision Table with different Federated learning Frameworks}
\end{figure*}

\vspace{10pt}

\section{Future Scope}

We propose a blockchain-based federated learning framework with advanced cryptographic techniques, opening new research avenues. Enhanced decentralized attribute-based encryption (DABE) algorithms could offer more secure, granular access control and reduce computational overhead for IoT devices. Improved differential privacy mechanisms might balance data leakage prevention with model accuracy. Optimizing homomorphic encryption and Secure Multi-Party Computation (SMPC) can boost aggregation efficiency and scalability for numerous nodes and complex models. The framework's adaptability could extend to smart healthcare and industrial IoT, with scalability tested in real-world settings. Automated tools for microservice management could ease deployment across diverse IoT environments.

\bibliographystyle{IEEEtran}

\vspace{12pt}

\end{document}